# Short-Term Power Prediction for Renewable Energy Using Hybrid Graph Convolutional Network and Long Short-Term Memory Approach

Wenlong Liao, Birgitte Bak-Jensen, Jayakrishnan Radhakrishna Pillai, Zhe Yang, and Kuangpu Liu

AAU Energy, Aalborg University

Aalborg, Denmark

(Contact: zya@energy.aau.dk)

*Abstract*—Accurate short-term solar and wind power predictions play an important role in the planning and operation of power systems. However, the short-term power prediction of renewable energy has always been considered a complex regression problem, owing to the fluctuation and intermittence of output powers and the law of dynamic change with time due to local weather conditions, i.e. spatio-temporal correlation. To capture the spatio-temporal features simultaneously, this paper proposes a new graph neural network-based short-term power forecasting approach, which combines the graph convolutional network (GCN) and long short-term memory (LSTM). Specifically, the GCN is employed to learn complex spatial correlations between adjacent renewable energies, and the LSTM is used to learn dynamic changes of power generation curves. The simulation results show that the proposed hybrid approach can model the spatio-temporal correlation of renewable energies, and its performance outperforms popular baselines on real-world datasets.

*Index Terms*--Renewable energy, power prediction, graph convolutional network, long short-term memory, deep learning.

## I. INTRODUCTION

With the increase of fossil energy consumption and environmental pollution, the effective use of renewable energy has become a hot topic. Wind farms and photovoltaic (PV) power plants are widely used and considered as very promising renewable energies whose permeability is gradually increasing in power systems [1], [2]. Although these renewable energies can bring positive environmental and economic benefits, their intermittent and fluctuating natures make it difficult to accurately forecast PV and wind powers, which pose challenges to the safe operation of power systems [3]. Therefore, it is necessary to develop accurate forecasting methods of PV and wind power generations to assist the safe operation and economic dispatch of power systems.

In respect of horizons, power prediction can be divided into several groups: long-term power prediction with year scales, medium-term power prediction with month scales, short-term power prediction with hour scales, and very short-term power prediction with minute scales. Generally, existing methods of short-term power prediction can be divided into three main categories: 1) Physical methods. They are usually developed on the basis of the lower atmosphere and sophisticated meteorological features, such as humidity, pressure, wind speed, and temperature [4]. Taking PV power prediction as an example, physical methods of the PV power prediction mainly include [5]: sky imagery methods, satellite imaging methods, and numerical weather prediction methods. Although these methods achieve outstanding performance, they require high computation costs, which seriously limit their applicability. 2) Statistical methods. They mainly include [6]: autoregressive moving average (ARMA), autoregressive (AR), and autoregressive integrated moving average (ARIMA), which extract features from lagged time series curves and meteorological factors, and then quantify the non-linear dynamic relationship between features and powers to obtain forecasts. Compared with physical methods, statistical methods are relatively cost-saving, because they do not require any expensive simulations beyond historical PV and wind powers after being trained offline. However, the forecasting performance of statistical methods usually drops with the increase of the time horizon [7]. 3) Artificial intelligence (AI)-based methods. The traditional artificial intelligence-based methods mainly include support vector machine (SVM) and multi-layer perceptron (MLP) [8], which ignore the spatio-temporal correlations, so that the change of meteorological features is not restricted by local weather conditions and they cannot predict the power generation curves accurately. Most recently AI-based approaches are proposed to capture strong correlations between renewable energies located in the vicinity, such as long short-term memory (LSTM) [9], convolutional neural network (CNN) [10], and hybrid model [11], which improve the forecasting accuracy of the target site by inputting feature information collected from neighboring sites to the models.

Further, the above-mentioned approaches are commonly used for datasets recorded from Euclidean domains (e.g., images and time series), while the input data of short-term power prediction considering the spatio-temporal correlation of renewable energies should be graph-structured data, which includes a correlation matrix between multiple renewable energies and their historical power generation curves. Existing methods have difficulties in dealing with the graph-structured data, so they simplify the graph-structured data into Euclidean data by ignoring correlation matrices, which limits the forecasting accuracy [12]. Recently, various graph neural networks defined in graph domains have shown convincing performance to handle the complex graph-structured data in different fields [13], such as traffic flow forecast, social recommendation, and drug discovery. The input data of short-term power prediction considering spatio-temporal correlation of renewable energies belongs to the graph-structured data, so graph neural networks should have the potential for short-term power prediction.

To improve forecasting accuracy, this paper proposes a new graph neural network-based short-term power forecasting approach, which combines the graph convolutional network (GCN) and LSTM to capture the spatio-temporal correlation simultaneously. The key contributions are as follows:

1) Multiple neighboring renewable energies are modeled as a graph, in which the adjacent matrix of nodes represents spatial dependencies.

2) A novel graph neural networks-based hybrid approach is proposed for short-term power prediction for renewable energies. Specifically, the GCN is used to capture the spatial dependence between multiple neighboring wind farms or PV plants, and the LSTM is employed to learn temporal features from the time series curves.

3) The influence of key parameters (e.g. the number of hidden layers, the size of the training epoch, and the choice of the optimizer) on the performance is analyzed, and the constructive suggestions of how to select these parameters in the proposed model are given.

The rest of the paper is organized as follows. Section II formulates the proposed method, and section III presents the process of the proposed method. Numerical experiments are performed and analyzed in section IV. Finally, section V summarizes the paper.

## II. METHODOLOGY

### A. Problem Definition

For short-term power prediction for renewable energies, the goal is to forecast the future power generation curves in a certain period of time given the historical data, such as historical power or meteorological features. Without loss of generality, the power generation curves of multiple wind farms and PV plants are used as an example of historical data in the experiment section.

Definition 1: Graph-structured data $G$. Specifically, the multiple renewable energies (e.g., Wind farms or PV plants) can be represented as an undirected graph $G=(V,E)$, where each renewable energy is treated as a node $v_i$. Specifically, $V = (v_1, v_2, ..., v_N)$ is a group of renewable energies. $N$ is the number of renewable energies, and $E$ is a set of edges between these renewable energies. Normally, a matrix $A \in R^{N \times N}$ is utilized to represent the connection relationship between nodes. For traffic flow forecast and social recommendation, the adjacency matrix only contains binary variables, which is equal to 1 if there is a link between nodes and 0 denotes there is no link [14]. By analogy, the adjacency matrix can be represented by the correlation matrix between multiple renewable energies. Specifically, this paper employs the absolute value of the Pearson correlation coefficient between nodes to represent the spatial correlation of neighboring wind farms or PV plants [2], and each element in the adjacency matrix is a real number, which ranges from 0 to 1.

Definition 2: Feature matrix $X^{N \times F}$. The historical data of renewable energies is considered as the attribute feature represented by $X^{N \times F}$. $F$ denotes the length of the historical time series. Again, attribute features of each node can be historical power generation curves or meteorological features, and the power generation curves of multiple wind farms and PV plants are used as an example in this paper.

In general, the problem of short-term power prediction for renewable energies can be regarded as learning a complicated neural network $f$, which projects a feature matrix and an adjacency matrix to the future power generation curves in a certain period of time:

$$X_t^{\text{all}} = [X_t^1, X_t^2, ..., X_t^N] \qquad (1)$$

$$\left[ X_{t+1}^l, ..., X_{t+k}^l \right] = f\left( A, \left( X_t^{\text{all}}, X_{t-1}^{\text{all}}, ..., X_{t-h}^{\text{all}} \right) \right), 1 \le l \le N \quad (2)$$

where $h$ is the length of historical power generation curves; $X_t^{\text{all}}$ is the set of power generation curves from multiple renewable energies at time $t$; $X_{t+1}^l$ is the predicted powers of the targeted renewable energy at time $t+1$; and $k$ is the length of future power generation curves needed to be predicted. Obviously, when $k$ is equal to 1, it is a one-step prediction, and when $k$ is greater than 1, it is a multi-step prediction.

The following section will explain how to use the proposed hybrid model to realize the short-term power prediction task. Specifically, the hybrid model includes two parts: a GCN and an LSTM. As shown in Fig.1, an adjacency matrix $A$ and historical power generation curves collected from past time $t$-$h$ to current time $t$ are input to the GCN, so as to obtain spatial features of multiple neighboring wind farms or PV plants. Then, the obtained spatial features are used as the input data to the LSTM, so as to capture temporal features by information transmission between renewable energies. Finally, the future power generation curves from time $t$+1 to $t$+$k$ are predicted through a dense layer with $k$ unit. The number of units in the dense layer is used to decide whether to make a one-step prediction or a multi-step prediction.

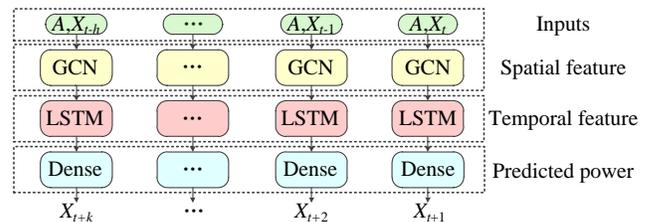

Figure 1. The framework of the proposed method.

## B. Modeling Spatial Correlation with GCN

Modeling the complicated spatial features is a key problem for the short-term power prediction of renewable energies. As shown in Fig. 2(a), despite the traditional CNN can obtain local spatial features of the red value of the red node along with its neighbors, it can only be used for the data defined in Euclidean domains, i.e., neighbors of each node are ordered and have a fixed size. To consider spatio-temporal correlations, the input data of short-term power prediction includes a correlation matrix between multiple renewable energies and their historical power generation curves, which belong to a graph rather than a 2-dimensional matrix. Different from the data in Euclidean domains, neighbors of each node are unordered and variable in size for the graph-structured data, as shown in Fig.2 (b). Therefore, the traditional CNN cannot make good use of the correlation matrix between multiple renewable energies and accurately capture spatial features. Recently, the traditional CNN in Euclidean domains has been generalized into the GCN in graph domains, which has shown outstanding performances in many fields [15], including text classification, fault diagnosis, and graph generation. To this end, the GCN is employed to model spatial features in this section.

The existing GCN mainly consists of two categories: spectral-based GCN and spatial-based GCN. Specifically, the former employs the Fourier transform to project the graph-structured data into the Fourier domains, and then the data is projected back to the graph domains after performing convolutional operations. In contrast, the latter directly defines convolutional operations on the graph domains by operating on spatially neighboring nodes. Both spectral-based GCN and spatial-based GCN are constantly developing and improving, it is difficult to say which one is better. Without loss of generality, a popular spectral-based GCN is used as an example to model the spatial features.

As shown in Fig. 3, The GCN can obtain the spatial correlation between the central renewable energy and its surrounding other power generation units by encoding the adjacency matrix and the feature matrix. The mathematical formula of each graph convolutional operation can be expressed as:

$$X_{\text{GCN}}^{(i+1)} = \sigma_{\text{GCN}}\left(\bar{A}X_{\text{GCN}}^{(i)}W_{\text{GCN}}^{(i)}\right), i = 1, 2, \dots M \quad (3)$$

$$\bar{A} = D^{-\frac{1}{2}}\hat{A}D^{-\frac{1}{2}}, \hat{A} = A + I, D_{ii} = \sum_j \hat{A}_{ij} \quad (4)$$

where $X_{\text{GCN}}^{(i)}$ is the feature matrix of the $i^{\text{th}}$ graph convolutional layer (The initial feature matrix includes $N$ historical power generation curves from past time $t$-$h$ to current time $t$); $M$ is the number of graph convolutional layers; $W_{\text{GCN}}^{(i)}$ is the weight matrix of the $i^{\text{th}}$ graph convolutional layer; $\sigma_{\text{GCN}}(\cdot)$ is the activation function of graph convolutional layers; $D$ is the diagonal node degree matrix of the adjacency matrix $A$; $I$ is an identity matrix; and $\hat{A}$ is a new form of the adjacency matrix with self-connection structure. Note that the output features of the last graph convolutional layer are used as the input features of the first LSTM layer.

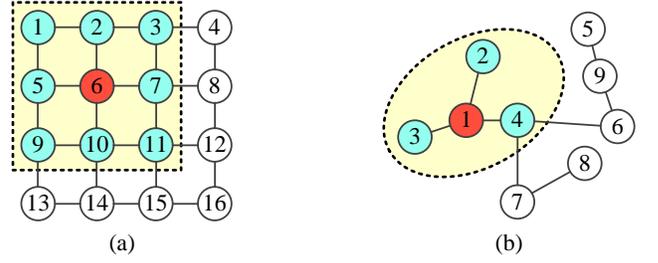

Figure 2. Euclidean convolution versus graph convolution. (a) Convolutional operations on Euclidean domains. (b) Convolutional operations on graph domains.

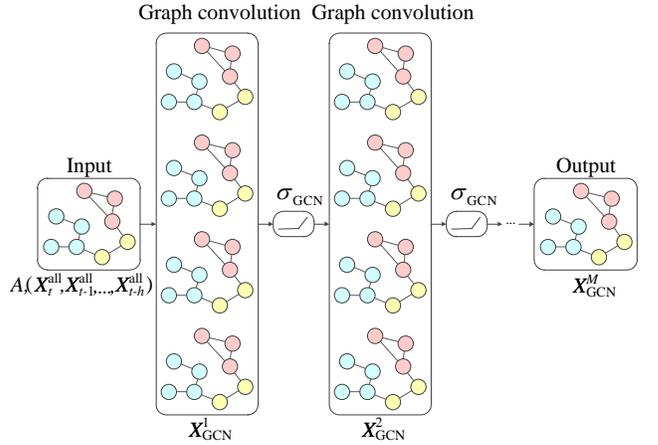

Figure 3. Graph convolutional operation on the graph-structured data.

## C. Modeling Temporal Correlation with LSTM

Modeling the complex temporal features is another key problem for short-term power prediction of renewable energies. So far, the recurrent neural network is one of the most widely used methods for short-term power prediction of time series. Nevertheless, the traditional recurrent neural network has gradient vanishing and exploding problems [16], which seriously limit its performance to learn long-term temporal correlations. To address these problems, the LSTM architecture was first proposed to memorize long-term dependence as much as possible in [17], and then further improved by adding an extra forget gate in [18]. At present, the LSTM has been the most popular recurrent neural network architecture and has shown convincing performance in many sequential tasks. Therefore, the LSTM is employed to model temporal features in this section.

As shown in Fig.4, there are three input features for each LSTM unit, which includes hidden state vector $H_{t-1}$ at time $t$-1, cell state vector $C_{t-1}$ at time $t$-1, and feature information $X_t$ at time $t$. Note that $H_t$ is considered the output of the LSTM layer. While modeling the feature information at the current moment, the LSTM still keeps the dynamic trend of historical power generation curves and shows the ability to capture temporal correlation. The output vectors of the LSTM unit can be obtained through and non-linear transformation and logical operation:

$$
\begin{aligned}
F_t &= \sigma_s \left( W_F X_t + U_F H_{t-1} + B_F \right) \\
I_t &= \sigma_s \left( W_I X_t + U_I H_{t-1} + B_I \right) \\
O_t &= \sigma_s \left( W_O X_t + U_O H_{t-1} + B_o \right) \\
\tilde{C}_t &= \sigma_g \left( W_C X_t + U_C H_{t-1} + B_C \right) \\
C_t &= F_t \circ C_{t-1} + I_t \circ \tilde{C}_t \\
H_t &= O_t \circ \sigma_g \left( C_t \right)
\end{aligned}
\tag{5}
$$

where $F_t$ is the activation vector of the forget gate; $I_t$ is the activation vector of the update gate; $O_t$ is the activation vector of the output gate; $\tilde{C}_t$ is the cell input activation vector; $\sigma_s$ is the sigmoid function; $\sigma_g$ is the hyperbolic tangent function; $W_F$ and $U_F$ are weight matrices of the forget gate; $W_I$ and $U_I$ are weight matrices of the update gate; $W_O$ and $U_O$ are weight matrices of the output gate; $W_C$ and $U_C$ are weight matrices of the cell state; $B_F$ is the bias vector of the forget gate; $B_I$ is the bias vector of the update gate; $B_o$ is the bias vector of the output gate; $B_C$ is the bias vector of the cell state; and $\circ$ is the Hadamard product. Note that the output features of the last LSTM layer are used as the input features of the dense layer.

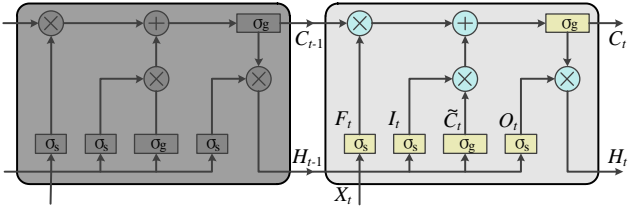

Figure 4.  The framework of the LSTM unit.

### D.  Short-term Power Prediction with Hybrid Form

Normally, the outputs of the last LSTM layer are fed to a dense layer which projects the intermediate LSTM outputs to future power generation curves from time $t+1$ to $t+k$. The mathematical formula of a dense layer can be expressed as:

$$
X_{\text{Dense}}^{(i+1)} = \sigma_{\text{Dense}} \left( X_{\text{Dense}}^{(i)} W_{\text{Dense}}^{(i)} + B_{\text{Dense}}^{(i)} \right)
\tag{6}
$$

where $X_{\text{Dense}}^{(i)}$ is the feature matrix of the $i^{\text{th}}$ dense layer; $W_{\text{Dense}}^{(i)}$ is the weight matrix of the $i^{\text{th}}$ dense layer; $B_{\text{Dense}}^{(i)}$ is the bias vectors of the $i^{\text{th}}$ dense layer; and $\sigma_{\text{Dense}} (\cdot)$ is the activation function of the dense layer.

### III.  Process of The Proposed Method

The process of short-term power prediction for renewable energies based on the proposed method is shown in Fig. 5, and the specific steps are as follows:

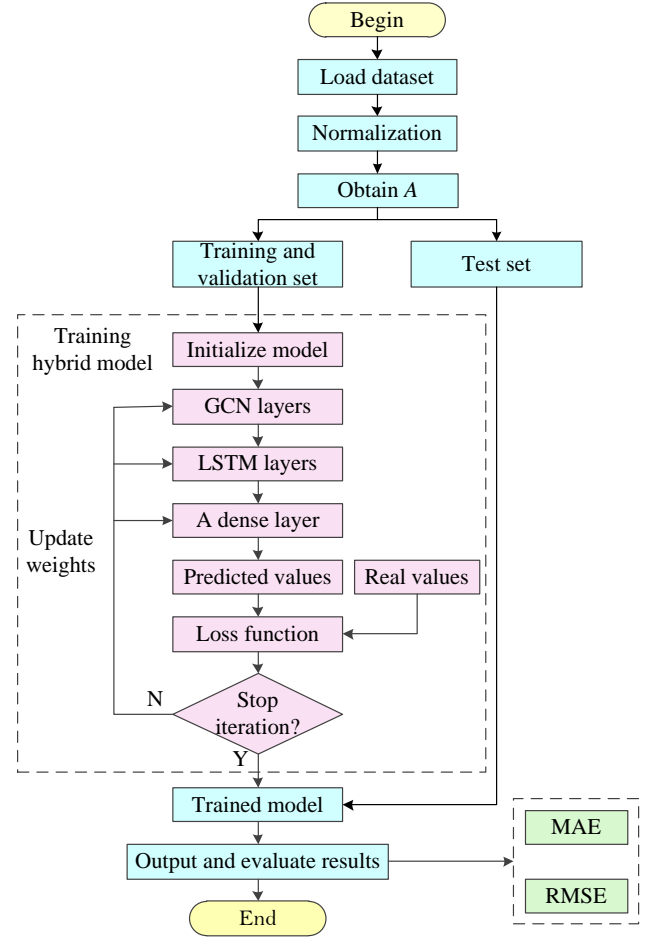

Figure 5.  Process of the proposed method.

1) Import and preprocess datasets. For short-term power prediction for renewable energies, the goal is to forecast the future power generation curves from time $t+1$ to $t+k$ given the historical data, such as historical power or meteorological features. Without loss of generality, the power generation curves of multiple neighboring wind farms or PV plants are used as an example of historical data in the experiment section. Then, the min-max normalization method is utilized to project the historical data into values that vary from 0 to 1. To account for the spatial correlation, the absolute value of the Pearson correlation coefficient between multiple neighboring renewable energies is employed to form an adjacency matrix $A$ for the GCN. Next, a part of samples are selected for the training set and validation set to fit the parameters of neural networks. The remaining samples are used to evaluate the performance of the pre-trained model.

2) Initializing parameters and train the model. To improve the performance of the proposed model, there is a need to explore the suitable structure and parameters before training the model. The parameters of the proposed model mainly include the numbers of middle layers (e.g., graph convolutional layers and LSTM layers), training epoch, and the selection of optimizer and its learning rate (LR).

Generally, the control variable method is utilized to adjust these parameters [15]. After initializing the parameters, the back-propagation algorithm is used to update the weights of the model by optimizing the loss function, such as mean absolute error (MAE). When the iteration ends, the pre-trained model is used to forecast future power generation curves.

3) Evaluate the performance of models. To evaluate the prediction performance of the proposed model and baselines for the test set, the MAE and the root mean square error (RMSE) are used to evaluate the difference between the real power $Y_t$ at time $t$ and the forecasting power $\hat{Y}_t$ at time $t$. The definitions of these two metrics are shown as:

$$MAE = \frac{1}{k} \sum_{t=1}^{k} \left| Y_t - \hat{Y}_t \right| \tag{7}$$

$$RMSE = \sqrt{\frac{1}{k} \sum_{t=1}^{k} \left( Y_t - \hat{Y}_t \right)^2} \tag{8}$$

For the RMSE and MAE, the smaller the value is, the stronger the performance of the model is.

## IV. CASE STUDY

### A. Data Description and Software Platform

To demonstrate the forecasting superiority of the proposed hybrid model based on GCN and LSTM, two datasets from the National Renewable Energy Laboratory (NREL) in the United States are employed [19], [20]. The first dataset includes 2190 wind power generation curves of 16 neighboring wind farms from January 1, 2007 to December 31, 2012, and the second dataset includes 1460 PV power generation curves of 9 neighboring plants from January 1, 2007 to December 31, 2010. The time resolutions of these power generation curves in two datasets are 10 minutes. The samples are divided into the training set, validation set, and test set according to seasons. In each season, the first 80% of the data is treated as the training set, followed by 10% of the data as the validation set, and the rest of the data as the test set.

The programming language is Python. The programs of different models for short-term power prediction are implemented in Spyder 4.1.5 with deep learning frameworks (e.g., Keras 2.3.1 and Tensorflow 2.1.0). The parameters of the computer are follows: Intel(R) Core(TM) i5-10210U, the processor base frequency is 1.60GHz, and the crucial laptop memory is 8 GB.

### B. Parameters Discussion

The hyper-parameters of the proposed hybrid model mainly include: past time length $h$, the numbers of middle layers, optimizer and its LR, and training epoch. In this paper, the control variable method is utilized to adjust these parameters through many experiments [15]. When one of the parameters is explored, the default values are used for the other parameters: The middle layer consists of 2 GCN layers and 2 LSTM layers. The optimizer is the Adam algorithm, and the LR is 0.001. The training epoch is 500.

As an example of predicting the wind powers for the next 1 hour, Fig. 6 shows the MAE of models with different past time length $h$.

Normally, one would expect to see a smooth U-shape, but the result appears to be that the random initialization of the parameters in the proposed model also dominates the choice of this hyper-parameter. Generally, the larger past time length $h$ is not the better. When $h$ is 6, the model has the smallest forecasting error.

Further, Table I shows the optimal past time length $h$ corresponding to different forecasting time length $k$. Normally, the optimal past time length $h$ varies from 6 to 12 for short-term wind power prediction. For short-term PV power prediction, the optimal past time length $h$ ranges from 8 to 118, and 114 can be considered as a good starting point for PV forecasts for the next 2 to 5 hours. Higher values or lower values may be fine for other PV power datasets. Note that the optimal past time length of the PV power is much larger than that of the wind power, which may be attributed to the strong diurnal trend of the PV power.

In order to explore the appropriate number of middle layers, Table II and Table III show the test set errors of models with different structures for short-term PV and wind power prediction of the next 1 hour.

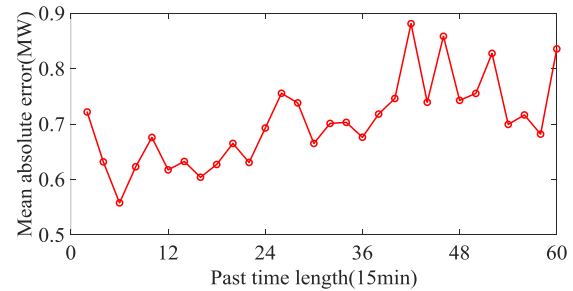

Figure 6. The MAE of models with different past time lengths.

TABLE I. THE OPTIMAL PAST TIME LENGTHS OF DIFFERENT TIME LENGTHS

| Forecasting time length (hour) | Past time length of wind power (10min) | Past time length of PV power (10min) |
|---|---|---|
| 0.5 | 6 | 8 |
| 1 | 6 | 80 |
| 1.5 | 12 | 118 |
| 2 | 6 | 98 |
| 2.5 | 12 | 118 |
| 3 | 6 | 116 |
| 3.5 | 6 | 114 |
| 4 | 8 | 118 |
| 4.5 | 12 | 104 |
| 5 | 8 | 114 |

TABLE II. THE MAE OF THE TEST SET FOR WIND POWER PREDICTION

| GCN and LSTM layers | | LSTM layers | | | |
|---|---|---|---|---|---|
| | | 1 | 2 | 3 | 4 |
| GCN layers | 1 | **0.82** | 0.86 | 1.01 | 1.43 |
| | 2 | 1.07 | 0.99 | 1.09 | 1.17 |
| | 3 | 1.06 | 1.20 | 1.17 | 1.19 |
| | 4 | 1.42 | 1.53 | 1.99 | 1.98 |

TABLE III. THE MAE OF THE TEST SET FOR PV POWER PREDICTION

| GCN and LSTM layers | LSTM layers | | | |
|---|---|---|---|---|
| | 1 | 2 | 3 | 4 |
| GCN layers 1 | 1.25 | 1.26 | 1.27 | 1.30 |
| 2 | 1.30 | 1.27 | 1.30 | 1.31 |
| 3 | 1.33 | 1.26 | 1.23 | **1.19** |
| 4 | 1.36 | 1.21 | 1.29 | 1.23 |

For short-term PV and wind power prediction, the number of middle layers in the hybrid model is not the more the better, since the capacity of the proposed hybrid model is way bigger than what is needed for short-term prediction, which results in a very high error on the test set, i.e., the over-fitting problem. Specifically, 1 GCN layer and 1 LSTM layer are suitable to form the middle layer of the hybrid model for the wind power dataset, and 3 GCN layers and 4 LSTM layers are suitable for the PV dataset. For other datasets, the number of middle layers can be adjusted according to forecasting errors of the validation set.

After initializing the structure of the hybrid model, it needs to select an appropriate optimizer to optimize the loss function. Mainstream optimizers include [21]: adaptive moment estimation (Adam), stochastic gradient descent (SGD), root mean square propagation (RMSProp), adaptive gradient descent algorithm (Adagrad), adaptive delta (Adadelta), adaptive moment estimation extension based on infinity norm (Adamax), and Nesterov-accelerated adaptive moment estimation (Nadam). To find a suitable LR, the Adam algorithm is regarded as an example. The models with different LRs are trained 500 epochs respectively, and their loss functions of the training set are visualized, as shown in Fig. 7.

When LR is greater than 0.1, the loss function of the hybrid model vibrates or even does not decrease. Conversely, too small LR requires more training epochs (e.g., LR=$1 \times 10^{-5}$), and may lead to never converge (e.g., LR=$1 \times 10^{-6}$). Generally, LR should not be too large or too small, and a good starting point can be range from $1 \times 10^{-4}$ to $1 \times 10^{-2}$. After setting a suitable LR, the training epochs can be initialized to 100, which is enough to ensure that the hybrid model has converged.

Further, the hybrid models with different optimizers are trained 30 times respectively, and the average loss functions of the training set are shown in Fig. 8.

From Fig. 8, it can be seen that the proposed hybrid model can obtain good performance when Adam, RMSprop, Adamax, and Nadam algorithms are used as optimizers. Specifically, Nadam algorithm is more suitable for short-term wind power prediction compared with other algorithms, while Adam is the optimal optimizer for short-term PV prediction. In addition, it is obvious that the loss functions of SGD, Adagrad, and Adadelta significantly larger than those of other algorithms, which indicates that these three optimizers are not suitable for short-term power prediction based on hybrid models.

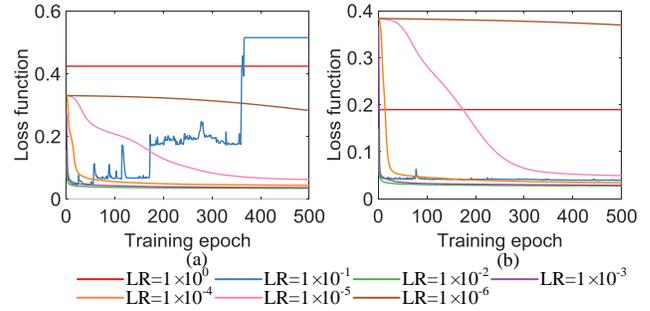

Figure 7. Loss functions of the hybrid model with different learning rates. (a) Wind farms. (b) PV plants.

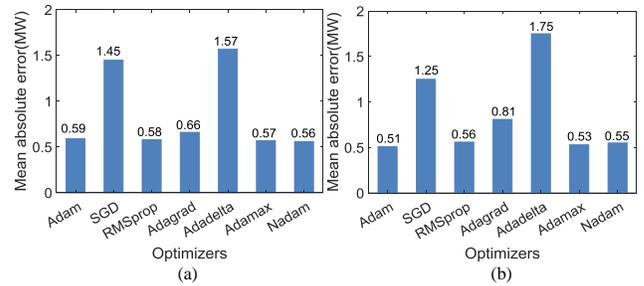

Figure 8. MSE of loss functions. (a) Wind farms. (b) PV plants.

### C. Comparison and Analysis with Popular Baselines

To illustrate the effectiveness of the proposed method, the hybrid model should be compared with popular baselines, such as MLP, LSTM, CNN, GCN, and the hybrid model of CNN and LSTM. Similarly, the control variable method is utilized to select suitable hyper-parameters through many experiments, as follows:

1) For MLP, the middle layer includes three dense layers, and the numbers of neurons are 30, 25, and 20, respectively [8]. 2) For LSTM, the middle layer includes 3 LSTM layers [9], and the sizes of units are 10, 15, and 10, respectively. 3) For CNN, the middle layer includes two 1-D convolutional (Conv1D) layers, two 1-D maximal pooling (MaxPooling1D) layers, a flatten layer, and a dense layer [10]. Specifically, the sizes of filters in Conv1D layers are 16 and 1 respectively. The size of the kernel in Conv1D layers is 1, and the pooling size in MaxPooling1D layers is 2. Besides, the unit of the dense layers is 1. 4) For GCN, 3 graph convolutional layers are used as the middle layer [15], and the sizes of filters are 20, 20, and 15, respectively. The output layer is a dense layer with 1 unit. 5) For the hybrid model of CNN and LSTM, it has a similar structure to the CNN [11]. Specifically, the hybrid model inserts an LSTM layer between the flatten layer and the dense layer of the CNN. The size of the unit in the LSTM layer is 10.

Besides, the training epochs, optimizers, and loss function of baselines are the same as the proposed hybrid model. Each method is independently repeated 30 times and the average MAE and RMSE of the test set are shown in Table IV.

TABLE IV. THE PREDICTION RESULTS OF THE PROPOSED HYBRID MODEL AND OTHER BASELINES

| Datasets | Forecasting time length | Indicators | MLP | GCN&LSTM | CNN&LSTM | CNN | LSTM | GCN |
|---|---|---|---|---|---|---|---|---|
| Wind farms | Forecasting time=1 hour | MAE(MW) | 1.11 | 0.82 | 1.02 | 0.99 | 1.03 | 0.90 |
| | | RMSE(MW) | 1.59 | 1.20 | 1.47 | 1.47 | 1.49 | 1.27 |
| | Forecasting time=2 hour | MAE(MW) | 1.93 | 1.23 | 1.42 | 1.63 | 1.44 | 1.45 |
| | | RMSE(MW) | 2.64 | 1.75 | 1.91 | 2.32 | 1.95 | 1.93 |
| | Forecasting time=3 hour | MAE(MW) | 2.35 | 1.80 | 1.90 | 2.22 | 1.91 | 1.95 |
| | | RMSE(MW) | 3.04 | 2.44 | 2.64 | 3.07 | 2.67 | 2.64 |
| | Forecasting time=4 hour | MAE(MW) | 2.76 | 2.18 | 2.47 | 2.62 | 2.50 | 2.36 |
| | | RMSE(MW) | 3.61 | 2.93 | 3.33 | 3.54 | 3.38 | 3.13 |
| | Forecasting time=5 hour | MAE(MW) | 3.16 | 2.69 | 3.06 | 3.15 | 3.07 | 2.74 |
| | | RMSE(MW) | 3.94 | 3.52 | 4.03 | 4.30 | 3.99 | 3.63 |
| Solar plants | Forecasting time=1 hour | MAE(MW) | 0.94 | 0.76 | 0.86 | 0.94 | 0.87 | 0.88 |
| | | RMSE(MW) | 1.97 | 1.57 | 1.72 | 1.84 | 1.73 | 1.81 |
| | Forecasting time=2 hour | MAE(MW) | 1.07 | 0.97 | 1.04 | 1.05 | 0.98 | 1.04 |
| | | RMSE(MW) | 2.25 | 1.99 | 2.14 | 2.08 | 2.07 | 2.08 |
| | Forecasting time=3 hour | MAE(MW) | 1.17 | 1.07 | 1.11 | 1.16 | 1.12 | 1.12 |
| | | RMSE(MW) | 2.38 | 2.21 | 2.28 | 2.39 | 2.31 | 2.36 |
| | Forecasting time=4 hour | MAE(MW) | 1.25 | 1.11 | 1.16 | 1.18 | 1.20 | 1.14 |
| | | RMSE(MW) | 2.60 | 2.29 | 2.34 | 2.48 | 2.45 | 2.33 |
| | Forecasting time=5 hour | MAE(MW) | 1.24 | 1.18 | 1.21 | 1.22 | 1.18 | 1.18 |
| | | RMSE(MW) | 2.55 | 2.38 | 2.43 | 2.48 | 2.49 | 2.42 |

The following conclusions can be drawn from Table IV: 1) A part of neural networks such as the proposed hybrid model and the LSTM, which focus on modeling the temporal features of power generation curves, generally show better forecasting performance than other baselines, such as the MLP. For example, for the 2-hours wind power prediction, the MAE of the proposed hybrid model and the LSTM are reduced by approximately 36.27 and 25.39% compared with the MLP, and the RMSE is approximately 33.71% and 26.13% lower than that of the MLP. This is because the traditional MLP has difficulty in handling non-stationary and complex time series curves. In addition, the forecasting precision of the GCN and CNN is not the highest, since they only account for the spatial features and ignore temporal features of power generation curves. 2) It is found that the MAE and RMSE of the proposed hybrid model are smaller than those based on a single model (e.g., LSTM or GCN), which indicates that the proposed hybrid model has the ability to accurately capture spatio-temporal features from power generation curves. For example, for the 1-hour PV power prediction, the MAE of the proposed hybrid model is reduced by approximately 13.64% compared with the GCN that only considers spatial features, and the RMSE is reduced by 13.26%. Compared with the LSTM which only considers temporal features, the MAE and RMSE of the proposed hybrid model are decreased by approximately 12.64% and 9.25% for the 1-hour PV power prediction. 3) Note that the hybrid model of the CNN and LSTM has a weaker performance than that of the proposed hybrid model of the GCN and LSTM, because the traditional CNN simplify the graph-structured data (i.e., the input data of short-term power prediction) into the Euclidean data by ignoring correlation matrices, which limits the forecasting accuracy. 4) In general, Table IV shows the forecasting results of the proposed hybrid model and popular baselines for 1 hour, 2 hours, 3 hours, 4 hours, and 5 hours on the wind power dataset and PV power dataset. It can be seen that the proposed hybrid model obtains the best performance under all evaluation indicators for all forecasting time horizons, proving the effectiveness of the hybrid model in spatio-temporal short-term power prediction of renewable energies.

## V. CONCLUSION

To improve the forecasting precision of short-term power predictions, a novel graph neural network-based hybrid approach is presented in this paper. After the simulation analysis on two real-world datasets, the following conclusions are obtained:

1) The optimal past time length $h$ varies from 6 to 12 for short-term wind power forecasts. For short-term PV power prediction, the optimal past time length $h$ ranges from 8 to 118, and 114 can be considered as a good starting point for PV forecasts for the next 2 to 5 hours.

2) The number of middle layers in the hybrid model is not the more the better, since the capacity of the proposed hybrid model is way bigger than what is needed for short-term prediction, which leads to the over-fitting problem. The proposed hybrid model can obtain good performance when Adam, RMSprop, Adamax, and Nadam algorithms are used as optimizers. Besides, LR should not be too large or too small, and a good starting point can be range from $1 \times 10^{-4}$ to $1 \times 10^{-2}$.

3) For the short-term PV and wind power prediction, the proposed hybrid model outperforms popular baselines (e.g., MLP, CNN, LSTM, GCN, and the hybrid model of CNN and LSTM) under different forecasting time horizons.

As a part of graph-structured data, the adjacency matrix of the proposed hybrid model is a fixed correlation matrix, which may be extended to a dynamic graph-structured data through spatial-temporal graph neural networks in future works. Also, the inputs to the proposed model do not involve the time of day and numerical weather prediction information, but they can easily be incorporated in the extension work.

## REFERENCES

[1] H. Zhang, Y. Liu, J. Yan, S. Han, L. Li, and Q. Long, "Improved deep mixture density network for regional wind power probabilistic forecasting," *IEEE Trans. Power Syst.*, vol. 35, pp. 2549-2560, Jul. 2020.

[2] L. Ge, W. Liao, S. Wang, B. Bak-Jensen, and J. R. Pillai, "Modeling daily load profiles of distribution network for scenario generation using

flow-based generative network," *IEEE Access*, vol. 8, pp. 77587-77597, Apr. 2020.

[3] Q. Zhao, W. Liao, S. Wang, and J. R. Pillai, "Robust voltage control considering uncertainties of renewable energies and loads via improved generative adversarial network," *J. Mod. Power Syst. Clean Energy.*, vol. 8, pp. 1104-1114, Nov. 2020.

[4] E. Pelikan, K. Eben, J. Resler, P. Jurus, P. Krc, M. Brabec, T. Brabec, and P. Musilek, "Wind power forecasting by an empirical model using NWP outputs," in *Proc. 9th Int. Conf. Environ. Elect. Eng.*, pp. 45-48.

[5] M. Sun, C. Feng, and J. Zhang, "Probabilistic solar power forecasting based on weather scenario generation," *Appl. Energy.*, vol. 266, pp. 1-12, May. 2020.

[6] S. Hu, Y. Xiang, H. Zhang, S. Xie, J. Li, C. Gu, W. Sun, and J. Liu, "Hybrid forecasting method for wind power integrating spatial correlation and corrected numerical weather prediction," *Appl. Energy.*, vol. 293, pp. 1-18, Apr. 2021.

[7] R. Huang, T. Huang, R. Gadh, and N. Li, "Solar generation prediction using the ARMA model in a laboratory-level micro-grid," in *Proc. 2012 IEEE 3rd Int. Conf. Smart Grid Commun.*, pp. 528-533.

[8] R. Deo, M. Ghorbani, S. Samadianfard, T. Maraseni, M. Bilgili, and M. Biazar, "Multi-layer perceptron hybrid model integrated with the firefly optimizer algorithm for windspeed prediction of target site using a limited set of neighboring reference station data," *Renew. Energy.*, vol. 116, pp. 309-323, Feb. 2018.

[9] F. Shahid, A. Zameer, A. Mehmood, and M. Raja, "A novel wavenets long short term memory paradigm for wind power prediction," *Appl. Energy.*, vol. 269, pp. 1-11, Jul. 2020.

[10] C. Huang and P. Kuo, "Multiple-input deep convolutional neural network model for short-term photovoltaic power forecasting," *IEEE Access.*, vol. 7, pp. 74822-74834, Jun. 2020.

[11] Z. Sun and M. Zhao, "Short-term wind power forecasting based on VMD decomposition, ConvLSTM networks and error analysis," *IEEE Access.*, vol. 8, pp. 134422-134434, Jul. 2020.

[12] M. Khodayar and J. Wang, "Spatio-temporal graph deep neural network for short-term wind speed forecasting," *IEEE Trans. Sustain. Energy.*, vol. 10, pp. 670-681, Apr. 2019.

[13] L. Zhao, Y. Song, C. Zhang, Y. Liu, P. Wang, T. Lin, M. Deng, and H. Li, "T-GCN: A temporal graph convolutional network for traffic prediction," *IEEE Trans. Intell. Transp. Syst.*, vol. 21, pp. 3848-3858, Sept. 2020.

[14] S. Georgousis, M. P. Kenning, and X. Xie, "Graph Deep Learning: State of the Art and Challenges," *IEEE Access.*, vol. 21, pp. 22106-22140, Jan. 2021.

[15] W. Liao, D. Yang, Y. Wang, and X. Ren, "Fault diagnosis of power transformers using graph convolutional network," *CSEE J. Power Energy Syst.*, vol. 7, pp. 241-249, Mar. 2021.

[16] R. Wang, C. Li, W. Fu, and G. Tang, "Deep learning method based on gated recurrent unit and variational mode decomposition for short-term wind power interval prediction," *IEEE Trans. Neural Netw. Learn. Syst.*, vol. 31, pp. 3814-3827, Oct. 2020.

[17] S. Hochreiter and J. Schmidhuber, "Long short-term memory," *Neural Comput.*, vol. 9, pp. 1735-1780, Dec. 1997.

[18] Y. Wang, W. Liao, and Y. Chang, "Gated recurrent unit network-based short-term photovoltaic forecasting," *Energies.*, vol. 11, pp. 1-14, Aug. 2018.

[19] C. Draxl, A. Clifton, B. Hodge, and J. McCaa, "The Wind Integration National Dataset (WIND) Toolkit," *Appl. Energy.*, vol. 151, pp. 355-366, Aug. 2015.

[20] Y. Zhang. (2012, Aug.). Solar integration national dataset toolkit. National Renewable Energy Laboratory., Golden, CO. [Online]. Available: https://www.nrel.gov/grid/sind-toolkit.html

[21] W. Liao, B. Bak-Jensen, J. R. Pillai, D. Yang, and Y. Wang. (2021, Jun.). Data-driven missing data imputation for wind farms using context encoder. *J. Mod. Power Syst. Clean Energy*. [Online]. pp. 1-13. Available: https://ieeexplore.ieee.org/document/9447776